\documentclass{ws-procs9x6}

\begin{document}

\title{Propensity matrix method for age dependent stochastic infectious disease models\footnote{\uppercase{T}his work was done in the framework of the \uppercase{H}ungarian \uppercase{N}ational \uppercase{D}evelopment, \uppercase{R}esearch, and \uppercase{I}nnovation (\uppercase{NKFIH}) \uppercase{F}und 2020-2.1.1-\uppercase{ED}-2020-00003.}}

\author{P. BOLDOG, N. BOGYA\footnote{\uppercase{W}ork partially
supported by grant \uppercase{NKFIH}-1279-2/2020 of the \uppercase{M}inistry for \uppercase{I}nnovation and \uppercase{T}echnology, \uppercase{H}ungary} and ZS. VIZI\footnote{\uppercase{W}ork partially
supported by grant \uppercase{F}unctional \uppercase{D}ifferential \uppercase{E}quations in \uppercase{M}athematical \uppercase{E}pidemiology \uppercase{N}ational \uppercase{R}esearch, \uppercase{D}evelopment and \uppercase{I}nnovation \uppercase{O}ffice \uppercase{NKFI FK} 124016}}

\address{University of Szeged, Bolyai Institute, \\
Aradi vértanúk tere 1, \\ 
6720, Szeged, Hungary\\ 
E-mail: boldogpeter@gmail.com}

\maketitle

\abstracts{Mathematical modeling is one of the key factors of the effective control of newly found infectious diseases, such as COVID-19. Our knowledge about the parameters and the course of the infection is highly limited in the beginning of the epidemic, hence computer implementation of the models have to be quick and flexible. The propensity matrix - update graph method we discuss in this paper serves as a convenient approach to efficiently implement age structured stochastic epidemic models. The code base we implemented for our forecasting work is also included in the attached GitHub repository\cite{vizigit}.}

\section{Introduction}

Quick and effective response to an emerging infection requires a large amount of information about the course of the epidemic (such as infection and recovery rate, latency period, etc.) and state of the art mathematical models. As it has been observed during the COVID-19 pandemic, the information at our disposal is highly limited, especially at the very beginning of the disease outbreak -- when we have to act quickly. Thus, our models, and therefore their computer implementations have to be modified drastically on a daily basis.

Experts from the field of mathematical epidemiology, such as Rost et al.\cite{Rost} and Barbarossa\cite{Barbarossa}, usually apply deterministic models to predict the spread of the disease as a first approximation. Nevertheless, realistic models (with several classes of infection, age structure or spatial patches) tend to get analytically intractable. Furthermore, deterministic models do not produce information about several important aspects of the epidemic, such as the variance of the state variables or the probability of certain events (like extinction) that are particularly important at the beginning of the outbreak. Stochastic modeling approach offers feasible alternatives for tackling the above mentioned problems. Although the governing equation (the so called stochastic master equation) is often mathematically intractable, numerical simulations of the corresponding Markov process and the average of the generated time series provides valuable information of several aspects of the epidemic.

Gillespie's stochastic simulation algorithm\cite{Gillespie-2} (SSA) was originally designed to produce exact realisations of the stochastic master equation in case of coupled chemical reaction systems. Since then, the scope of the SSA has been extended by extensive research on this field\cite{perspective} in the last 40 years to stochastically simulate other chemical or biological phenomena such as diffusion processes, migration of cells and animals, or disease spread. Gillespie's algorithm yields a convenient method of simulation to gather information of an epidemic in the stochastic approach.

During the COVID-19 pandemic, one important lesson was that, quick implementation of complicated stochastic epidemic models with several age groups is just as crucial as running time or resource efficacy. The aim of the present paper is to provide a framework that enables researchers to quickly build and modify stochastic epidemic models with age structure.

The structure of the paper is as follows: in the remaining subsections of the Introduction, we introduce Gillespie's first reaction method and provide a Python code for the case of the SIR model. In Section 2, we introduce our propensity matrix approach and extend our model implementation with age structure. Then, we demonstrate the flexibility of our approach by generalizing the model with other state variables to investigate an epidemiologically more feasible age-stratified model containing the E (exposed) and the D (dead) classes additionally. In Section 3, we introduce the update graph to make the algorithm much more efficient in terms of computation. Then in Section 4, we demonstrate the strength of stochastic epidemic models via some experiments with a hypothetical population. Finally, we summarize and discuss our results in Section 5.

\subsection{Gillespie's stochastic simulation algorithms}\label{sec:gil-method}

Considering a group of chemical species ($S_1,S_2,\ldots,S_n$) reacting in a coupled system with reaction channels ($R_1, R_2,\ldots,R_m$) in a well mixed environment, Gillespie's stochastic simulation algorithms are based on answering two questions\cite{Gillespie-2}: 
\begin{romanlist}
	\item What kind of reaction ($R_1, R_2,\ldots,R_m$) will the next reaction be?\label{Q1}
	\item In what time ($\tau$) will it occur?\label{Q2}
\end{romanlist}
Gillespie formulated two distinct, but mathematically identical versions of the SSA: the so called 'direct method' (DM) and the 'first-reaction method' (FRM). The main difference is in the way of deciding which reaction channel to fire. The \textit{first-reaction method} generates $m$ random time values $(\tau_1,\ldots,\tau_m)$ for the $m$ possible reactions from the corresponding exponential distributions and selects the reaction channel with the least time to fire. Alternatively, the \textit{direct method} only requires two random numbers (one from a uniform distribution and one from an exponential distribution). 
Considering the fact that if there are more than 2 possible reactions then the FRM requires more computation and memory than the DM, moreover drawing $m$ random numbers from exponential distributions requires calculating $m$ logarithms, thus the computational cost of the \textit{first-reaction method} is even greater\cite{Gillespie-1}. As we are about to work with robust systems of reactions, we only review and make use of the \textit{direct method} in the following.

\subsubsection*{The direct method algorithm}

The direct method SSA creates stochastic realizations of the corresponding Markov chain that is continuous in time and discrete in state variables. Technically, starting from the vector of initial values $(X_1(0),\dots,X_n(0))\in\mathbb{N}_0^n$, the time series of the state variables $(X_1(t),\dots,X_n(t))\in\mathbb{N}_0^n$ is generated by constantly updating the state variables in properly generated subsequent times ($\tau$) according to the answers to question (\ref{Q1}) and (\ref{Q2}).

To this end, the reaction probability density function, $P(\tau, \mu)d\tau$, is defined\cite{Gillespie-1}. That is, at time $t$, the next reaction in the reaction chamber will occur in the differential time interval $(t + \tau, t + \tau + d \tau)$ with the probability $P(\tau, \mu)d\tau$, and will be an $R_\mu$ reaction ($\mu\in {1,\ldots,m}$).

It can be shown that\cite{Gillespie-1}, with the procedure called conditioning, the two variable density function $P(\tau,\mu)$ can be written as the product of two one-variable probability density functions:
\begin{equation}\label{eq:cond}
	P(\tau,\mu) = P_1(\tau) \cdot P_2(\mu|\tau).
\end{equation}
In particular, from the paper of Gillespie\cite{Gillespie-1}, it turns out to be $$P(\tau,\mu) = a_\mu \cdot e^{-\tau\cdot a},$$ where the symbol $a_\mu$ stands for the so called propensity function that characterizes reaction $R_\mu$, and may depend on the quality, and actual number of the reactants or the environment, etc. For convenience, we use the notation $a=\sum_{\mu=1}^m a_\mu$ for the sum of the propensity functions.

Let $P_1(\tau)$ be the probability that the next reaction will occur between times $t+\tau$ and $t+\tau+ d\tau$, independent of which reaction it might be. Similarly, $P_2(\mu|\tau)$ is the probability that the next reaction will be the $R_\mu$ reaction, given that the next reaction occurs at $t+\tau$. The probability $P_1(\tau)$ is obtained by summing the reaction probability density function over all possible $\mu$ values:
\begin{equation}\label{eq:P1}
	P_1(\tau)=\sum_{\mu=1}^m P(\tau,\mu)=a\cdot e^{-\tau\cdot a}.
\end{equation}
Substituting $P_1(\tau)$ into (\ref{eq:cond}) and solving for $P_2(\mu|\tau)$ we obtain
\begin{equation}\label{eq:P2}
	P_2(\mu|\tau)=\frac{a_\mu}{a}.
\end{equation}
It is clear that $\int_0^\infty P_1(\tau)d\tau=\int_0^\infty a\cdot e^{-\tau\cdot a}=1$. Moreover, $\sum_{\mu=1}^{m}P_2(\mu|\tau)=\sum_{\mu=1}^{m}\frac{a_\mu}{a}=1.$

With these notations and probability distributions, the direct method SSA can be described as follows.
\begin{enumerate}
	\item \textbf{Initialization:}
	\begin{alphlist}
		\item set $t\leftarrow0$,
		\item set initial values $(X_1(0),\dots,X_n(0))\in\mathbb{N}_0^n$, 
		\item prescribe halting conditions $C_H$.
	\end{alphlist}
	\item \textbf{Calculate propensity functions} $a_\mu$ for all $\mu\in{1,2,\ldots,m}$.
	\item \textbf{Decide when the next reaction will occur:} choose $\tau$ according to eq.~(\ref{eq:P1}),
	\begin{alphlist}
		\item choose $r_1$ from $(0,1)$ with a uniform distribution,
		\item obtain $\tau=(1/a)\ln(1/r_1)$.
	\end{alphlist}
	\item \textbf{Decide which reaction occurs:} choose $\mu$ according to eq.~(\ref{eq:P2}),
	\begin{alphlist}
		\item choose $r_2$ from $(0,1)$ with a uniform distribution,
		\item take $\mu$ to be the integer for which $$\sum_{j=1}^{\mu-1}a_j<r_2a\leq \sum_{j=1}^{\mu}a_j$$.
	\end{alphlist}
	\item \textbf{Update time and state variables:}
	\begin{alphlist}
		\item calculate change $(\Delta X_1,\dots,\Delta X_n)\in \mathbb{Z}^n$ in number of reactants according to $\mu$,
		\item change the number of molecules according to $$(\!X_1\!(t+\tau),\!\dots\!,X_n\!(t+\tau)\!)\!=\!\!(\!X_1(t)\!,\!\dots\!,\!X_n(t)\!)\!+\!(\!\Delta\! X_1\!,\!\dots\!,\!\Delta X_n\!),$$
		\item set $t\leftarrow t+\tau$.
	\end{alphlist}
	
	\item \textbf{Halt if $C_H = True$ else continue the process with Step (2).}
\end{enumerate}

\subsection{Gillespie's algorithm for the SIR model}

The well known deterministic SIR model is a system of ordinary differential equations (\ref{eq:SIR-ODE}), and it is one of the simplest models to capture the spread of an infection in a population without demography. It separates the population into three classes: $S(t)$ - the number of susceptible, $I(t)$ - the number of  infected, and $R(t)$ - the number of recovered individuals in the population at time $t$, the symbol $()'$ stands for the usual time derivative. The model defines two reactions that we will call events from now on. Susceptible individuals \textit{get infected} with rate $\beta$ by making effective contact with infected individuals in the population. Infected individuals \textit{recover} at a rate $\gamma$, and the per capita contact between $S$ and $I$ is $\frac{SI}{N}$. We consider a constant population size $N=S(t)+I(t)+R(t)$ and the following rate of change in the classes:

\begin{center}
	\begin{minipage}{.4\textwidth}
		\includegraphics[clip,trim=60 80 60 80,scale=0.4]{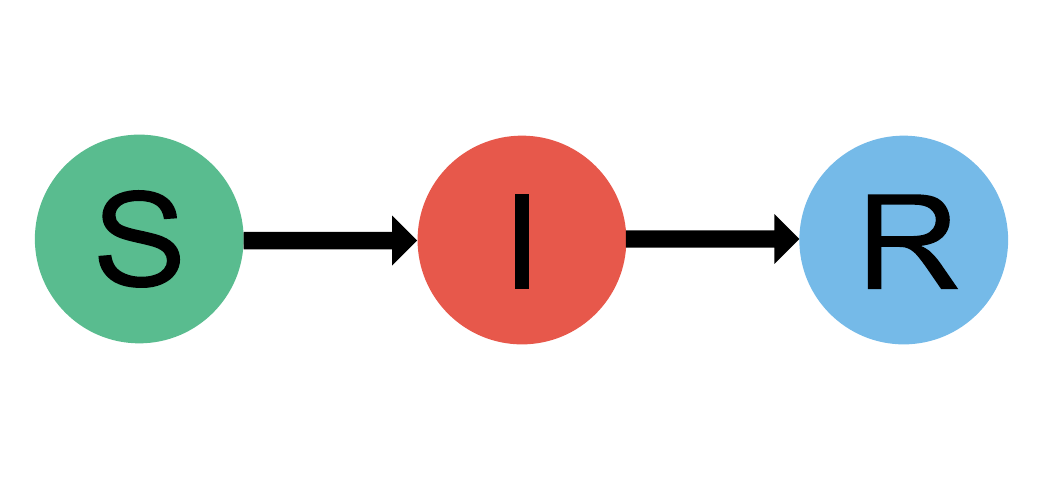}
	\end{minipage}\begin{minipage}{.4\textwidth}
		\begin{equation}\label{eq:SIR-ODE}
			\begin{cases}
				S'&=-\beta \frac{SI}{N}\\
				I'&=\beta \frac{SI}{N}-\gamma I\\
				R'&=\gamma I.
			\end{cases}
		\end{equation}
	\end{minipage}
\end{center}

In this section, by following the work of Gillespie\cite{Gillespie-2} and Allen\cite{Allen-primer}, we apply the \textit{direct method} SSA to the SIR model \eqref{eq:SIR-ODE}. The code we implemented can be found in our GitHub repository\cite{vizigit}. 
It is important to note that, as we prepare to handle robust models in a flexible way, we will not follow the common practice of reducing the state variables with decoupling \eqref{eq:SIR-ODE} by using the fact that $N$ is constant in time.

In the stochastic SIR model, state variables $S(t), I(t), R(t) \in\{0,1,\ldots,N\}$, and $t\in[0,\infty)$. In Table (\ref{tab:SIR}), we summarise the possible events with the corresponding transitions, change in state variables, and we also define the belonging propensity functions and probabilities. The intervent time $\tau$ is generated from the exponential distribution (\ref{eq:P1}) with parameter $1/a$, where $a=\beta\frac{SI}{N}+\gamma I$ and the probability distribution of the possible events is $p_1=a_1/a$ and $p_2=a_2/a$ (with propensities $a_1=\beta\frac{SI}{N}, a_2=\gamma I$).

\begin{table}
	\tbl{The table shows the possible events in the SSA in case of the SIR epidemic model.}
	{\footnotesize
		\begin{tabular}{@{}lcccc@{}}
			\hline
			{} &{} &{} &{} &{}\\[-1.5ex]
			Event & Transition & Change ($\Delta S,\Delta I, \Delta R$)  & Propensity & Probability \\[1ex]
			\hline
			{} &{} &{} &{} &{}\\[-1.5ex]
			Infection & $S\rightarrow I$ &$(-1,+1,0)$& $a_1=\beta\frac{SI}{N}$ & $p_1=\frac{a_1}{a}$ \\[1ex]
			Recovery & $I\rightarrow R$ &$(0,-1,+1)$& $a_2=\gamma I$ & $p_2=\frac{a_2}{a}$ \\[1ex]
			\hline
		\end{tabular}\label{tab:SIR}}
\end{table}

\section{The propensity matrix method for age dependent stochastic epidemic models}\label{sec:prop_mtx}

In order to extend the SSA algorithm to handle the required age structure, we make the following definitions and assumptions:
\begin{itemize}
    \item assume a population of $N$ individuals,
	\item the population is divided into $l$ \textit{classes} ($l=3$ in the SIR model) according to the characteristics of the epidemic: $X_1,\dots, X_l$,
	\item the population is stratified into $n$ \textit{age groups}, 
	\item this partition of the population leads us to the $n\cdot l$ \textit{compartments} (that we will call \textit{state variables} as well) at time $t$:
	$X_1^1(t),X_2^1(t),\dots,X_{l-1}^n(t), X_l^n(t)$,
	\item an \textit{event} (or reaction) during the simulation corresponds to a \textit{transition} of an individual from one compartment to another in the same age group,
	\item we consider the epidemic in such a short time scale when aging is not significant, thus we assume there is no transition between age groups,
	\item we assume that there is no demography (birth or natural death) during this short time -- if death is incorporated in the model,  it only occurs due to the infection,
	\item the same transitions occur in every age group -- possibly with zero rate.
\end{itemize}

Now, we define the propensity matrices as follows: for age group $i$, we define \textit{propensity matrix} $M^i$ with index $i$ in the following way: element of the propensity matrix $M^i(j,k)= a_{j,k}^i$ represents the propensity function corresponding to the transition of an individual from compartment $X_j\rightarrow X_k$ at age group $i$.

During the simulation, we calculate the propensity matrix for every age group. Since, we observe the same transitions in every age group, we can outsource the repeating calculations into $n$ cycles,  obtaining a cleaner and more compact code.

Then we choose the intervent time $\tau$ from distribution (\ref{eq:P1}), where $a$ is the sum of all propensities over all propensity matrices:
\begin{equation}\label{eq:a}
	a=\sum_{g=1}^{n}\sum_{f=1}^l\sum_{h=1}^l a_{f,h}^g,
\end{equation}
and choose the next event from distribution
\begin{equation}\label{eq:prop_mtx_event}
	P_2\left((\text{transition }j \rightarrow k \text{ at age group }i )|\tau\right)=p_{j,k}^i=\frac{a_{j,k}^i}{a}.
\end{equation} 
Basically, one can obtain a transition from distribution (\ref{eq:prop_mtx_event}) by choosing a random real $r\in(0,1)$ with uniform distribution and summing the propensities of all transitions over all age groups, then choosing transition $X_j\rightarrow X_k$ at age group $i$ whenever $$\sum_{g=1}^{n}\sum_{f=1}^l\sum_{h=1}^l a_{f,h}^g < r\cdot a\leq \sum_{g=1}^{i}\sum_{f=1}^j\sum_{h=1}^k a_{f,h}^g.$$
Eventually, we have to update time: $t\leftarrow t+\tau$ and the state variables with the change vector $(\Delta X_1^1,\ldots,\Delta X_l^n)$, where only $\Delta X_j^i=-1$ and $\Delta X_k^i=+1$ are nonzero elements: $$(X_1^1(t+\tau),\ldots,X_l^n(t+\tau))=(X_1^1(t),\ldots,X_l^n(t))+(\Delta X_1^1,\ldots,\Delta X_l^n).$$
With these assumptions and notations we are prepared to present the propensity matrix algorithm that derives the time evolution of the state variables. 
\subsection*{Algorithm propensity matrix method}

\begin{enumerate}
	\setcounter{enumi}{-1}
	\item Fix the order of the state variables: $(X_1^1,\ldots,X_l^n)$.
	\item \textbf{Initialize:} Set $t\leftarrow 0$, initial values $(X_1^{1}(0),\ldots,X_l^{n}(0))$, and halting condition $C_H$.
	\item \textbf{Calculate propensity matrices} $M_i$ for all age groups $i\in\{1,\ldots,n\}$.
	\item \textbf{Choose intervent time} $\tau$ from $P(\tau)=a\cdot e^{-\tau\cdot a}$ where $a$ comes from \eqref{eq:a},
	\item \textbf{Choose the next reaction} from the distribution  $p_{j,k}^i=\frac{a_{j,k}^i}{a}$ \eqref{eq:prop_mtx_event},
	\item \textbf{Update state variables and time.}
	\item \textbf{Halt} if $C_H=True$ else \textbf{continue} with step 2.
\end{enumerate}

Propensity matrices of epidemic models are sparse, since, from one compartment, there are usually only one or very few transitions to other compartments. We also emphasise that this data structure provides a convenient way to easily modify our model by including new transitions (like waning immunity), or deleting transitions between age groups. Also, by extending the number of state variables, we can further introduce new classes (like exposed, latent, hospitalized, dead or recovered). In the following, we present the propensity matrices of two epidemic models.

\subsection{The age structured SIR model with waning immunity}

When we extend model \eqref{eq:SIR-ODE} with age structure we have to consider the effect of every infected age group on the susceptible population of age group $i$, furthermore to include waning immunity to (\ref{eq:SIR-ODE}) we consider the $R_i\rightarrow S_i$ transitions with age dependent rate $\omega_i$, in the general case. Naturally, with $\omega_i=0$ for $i\in\{1,\ldots,n\}$ we get back to the case where waning immunity is not applied in the process.  Hence, with age dependent transmission rates $\beta_i$, recovery rates $\gamma_i$ and waning rates $\omega_i$ we obtain the system of ODEs for age group $i$:
\begin{equation}
	\begin{cases}
		\displaystyle S_i{'}=-\frac{1}{N} S_i\cdot \sum_{f=1}^n \beta_f I_f+\omega_i S_i\\
		\displaystyle I_i{'}=\frac{1}{N} S_i\cdot \sum_{f=1}^n \beta_f I_f-\gamma I_i\\
		\displaystyle R_i{'}=\gamma I_i-\omega_i S_i
	\end{cases}
\end{equation}

\begin{figure} 
	\centering
	\begin{minipage}{0.45\linewidth}
		\centering
		$\begin{cases}
			\displaystyle a_{SI}^i=\frac{1}{N} S_i\cdot \sum_{f=1}^n \beta_f I_f\\
			\displaystyle a_{SR}^i=\alpha_i S_i\\
			\displaystyle a_{IR}^i=\gamma_i I_i
		\end{cases}$
	\end{minipage}\begin{minipage}{0.45\linewidth}
		\centering
		\includegraphics[clip,trim=150 80 160 60,scale=0.3]{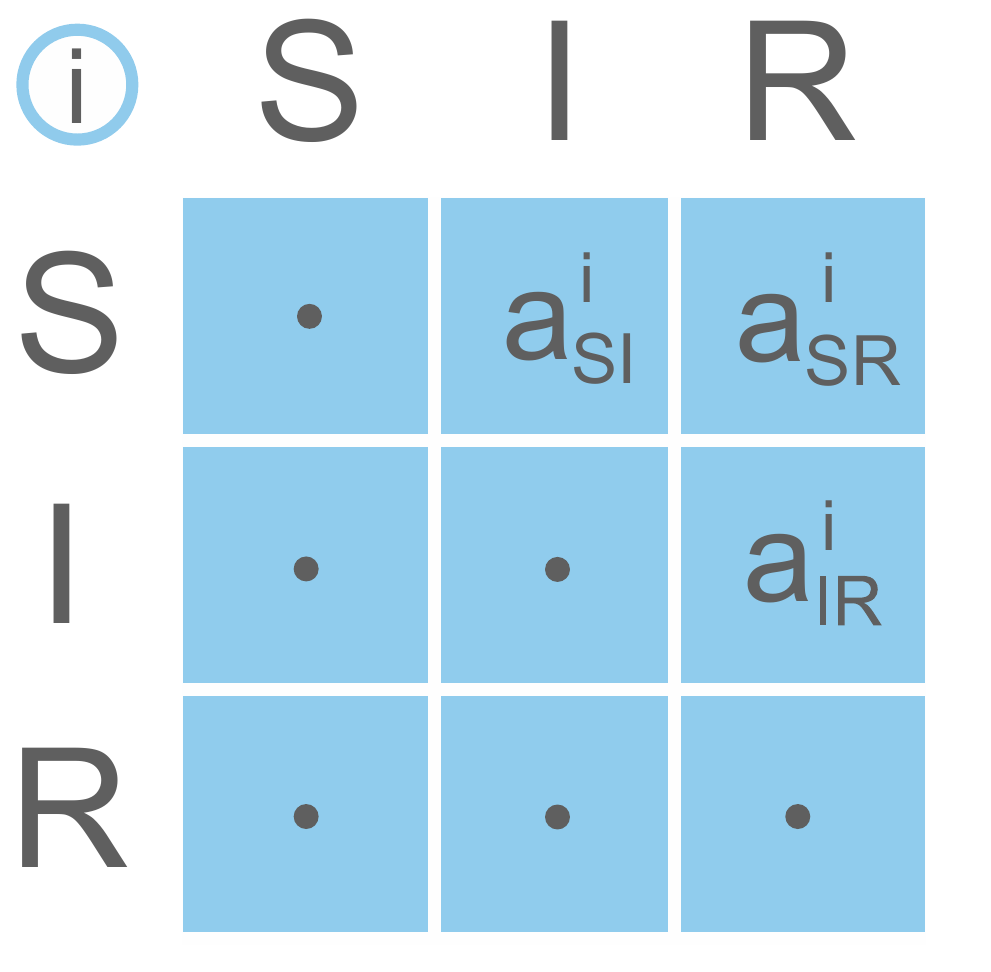}
	\end{minipage}  
	\caption{Propensity matrix and the corresponding propensities of the SIR model\label{fig:Prop-SIR-vac}}
\end{figure}

In the stochastic simulation formulation, let the order of state variables be $(X_1^i=S_i, X_2^i=I_i,X_3^i=R_i)$, thus let and with this notation, $a_{j k}^i$ is the propensity of transition from compartment $X_j$ to $X_k$ at age group $i$. Thus, the possible transitions in age group $i$ (and the belonging propensities) in this model are \textit{infection} ($a_{SI}^i$), \textit{waning immunity} ($a_{SR}^i$) and \textit{recovery} ($a_{IR}^i$). The propensity matrix and the corresponding propensities can be seen in Figure~\ref{fig:Prop-SIR-vac}, where '$\bullet$' stand for a zero element of the matrix. Python code for this model is included in the repository\cite{vizigit}.

\subsection{SEIRD model with age structure and waning immunity}\label{sec:SEIRD}

One strength of the propensity matrix technique is the flexibility in terms of computer implementation. For instance several disease models require an exposed class E as right after infection susceptibles usually do not show symptoms during the latency period of length $1/\epsilon$. Thus with rate $\epsilon$ people move from class $E_i$ to class $I_i$. The corresponding propensity is $a_{EI}^i=\epsilon E_i$. We can further add class D that counts the death cases caused by the infection - with age dependent mortality rate $p_i\gamma$ people move from class $I_i$ to $D_i$ with corresponding propensity, $a_{ID}^i=p_i\gamma I_i$. We also assume that recovered people loose immunity against the disease with rate $\omega$, thus, the belonging propensity is $a^i_{RS}=\omega R_i$.

\begin{figure}
	\centering
	\begin{minipage}{0.28\linewidth}
		\includegraphics[clip,trim=12 60 12 60,scale=0.23]{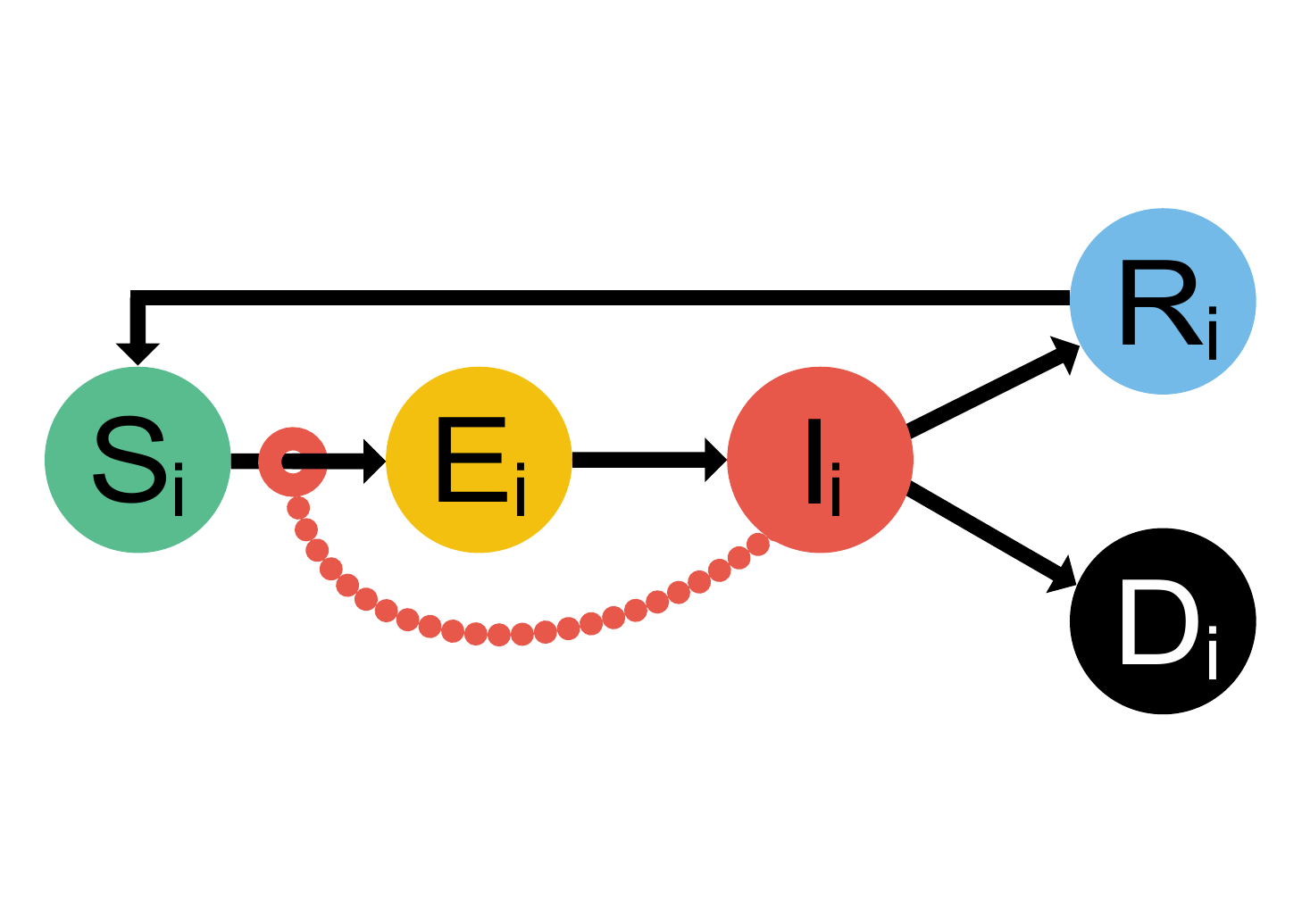}
	\end{minipage}
	\begin{minipage}{0.5\linewidth}
		\centering
		{\footnotesize $\begin{cases}
			S_i{'}=-\frac{1}{N_i} S_i\cdot \sum_{f=1}^n \beta_f I_f+\omega R_i \\
			E_i{'}=\frac{1}{N_i} S_i\cdot \sum_{f=1}^n \beta_f I_f-\epsilon E_i\\
			I_i{'}=\epsilon E_i-\gamma I_i\\
			R_i{'}=(1-p_i)\gamma I_i-\omega R_i\\
			D_i{'}=p_i \gamma I_i
		\end{cases}$}
	\end{minipage}\begin{minipage}{0.2\linewidth}
		\includegraphics[clip,trim=175 90 180 90,scale=0.28]{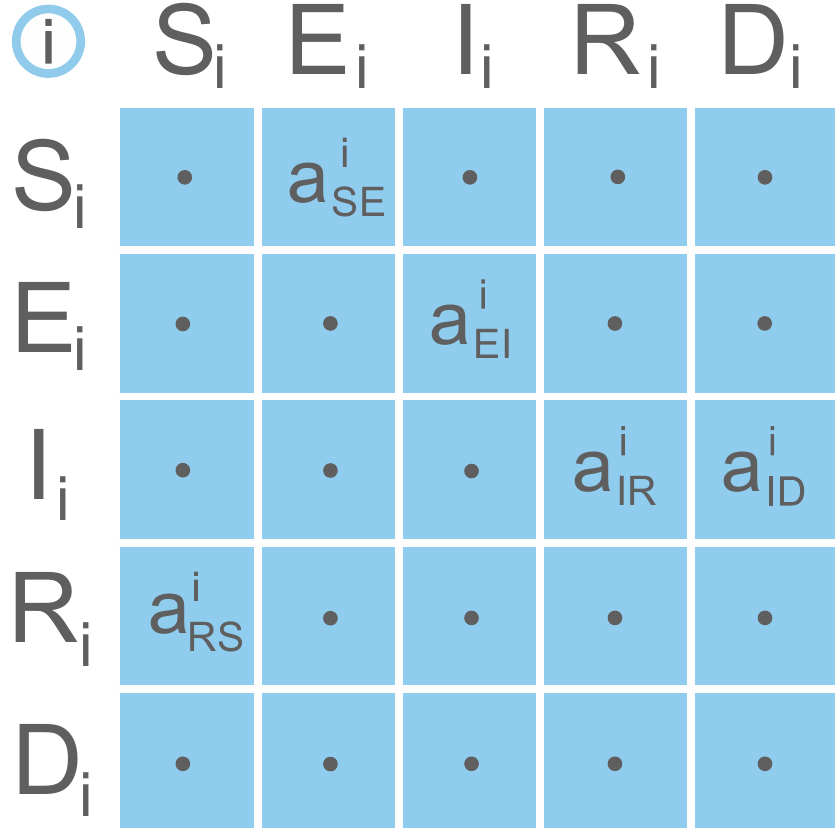}
	\end{minipage}
	\caption{\textbf{The SEIRD model with age structure and waning immunity.} \newline Possible transitions with the belonging propensities are: infection ($a_{SE}^i=\frac{1}{N_i} S_i\cdot \sum_{f=1}^n \beta_f I_f$), going infectious from exposed ($a_{EI}^i=\epsilon E_i$), recovery ($a_{IR}^i=(1-p_i)\gamma I_i$), death due to infection  ($a_{ID}^i=p_i\gamma I_i$) and loss of immunity  ($a_{RS}^i=\omega R_i$).\label{fig:SEIRD}}
\end{figure}

Let the order of state variables at age group $i$ be $X_1^i=S_i, X_2^i=E_i, X_3^i=I_i, X_4^i=R_i,  X_5^i=D_i$ and $a_{j k}^i$ stands for the propensity of transition from compartment $j$ to $k$ at age group $i$. For flowchart of the process and the governing ODE model with the propensity matrix and the remaining propensities see Figure~\ref{fig:SEIRD}.

\section{Updating the propensities}\label{sec:update}

Gillespie\cite{Gillespie-2} suggests that "...it is necessary to recalculate only those quantities $a_\nu$, corresponding to reactions $R_\nu$, whose reactant population levels were just altered..." in the reaction selection step. 

Probably the most well known attempt to get around this concern belongs to Gibson and Bruck\cite{Gibson-Bruck}. Their \textit{Next Reaction Method} (NRM) may be regarded as an extension of Gillespie's original first reaction method. They define the so called reaction dependency graph that contains information about which propensity function ($a_\nu$) needs to be updated according to the chosen reaction in the selection step. However, in the detailed comparison of the DM, FRM and NRM by Cao et al.\cite{Cao-Li-Petzold} it is found that "even with the best data structure, the NRM is less efficient than the DM except for a very specialized class of problems". As Scvehm\cite{Schwehm} points out, this is mostly due to the fact that in case of the Gibson-Bruck method "the simulator engine spends most of its execution time for maintaining the priority queue of the tentative reaction times".

In this section we provide a dependency graph like method to make use of the fact that upon a transition event mostly only a small number of propensity functions have to be recalculated. For instance in case of the age structured SEIRD model in Section \ref{sec:SEIRD} the waning immunity event at age group $i$ only changes state variable $R_i$ and $S_i$ hence it only influences propensity function $a^i_{RS}$ and $a^i_{SE}$. All other propensities in age group $i$ and every other propensities in the other age groups remain unchanged, that is $3+(n-1)\cdot5=5n-2$ number of events in case of $n$ age groups.

In our methodology, in every iteration step after choosing the transition from distribution (\ref{eq:prop_mtx_event}), we selectively update the propensity values based on this transition. We do this with the help of a suitable data structure that is a directed bipartite graph, called the \textit{Update Graph} (UG) and is defined the following way: let $C$ be the set of nodes associated with the \textit{classes} (cf. Sec. \ref{sec:prop_mtx}) and let $T$ be the other set of nodes representing the possible \textit{transitions} between the classes. The edge $c\rightarrow t_{X_j, X_k}$ (where $t_{X_j, X_k}$ is the  transition from $X_j$ to $X_k$) exists only if the class of a modified state variable $c\in C$ updates the propensity of transition $t_{X_j, X_k}\in T$. Also edge $t_{X_j,  X_k}\rightarrow c$ exists only if $c=X_j$ or $c=X_k$. For convenience we note that from any node $t_{X_j, X_k}\in T$ there are always two edges pointing out, one to class $X_j$ and the other one to class $X_k$. 

However from class $c\in C$ there may be several edges pointing to different transitions, for instance in the $SEIRD$ model infection, recovery and death are all depending on the number of $I$ individuals in a given age group. 
We also point out that (in case of the $SEIRD$ model) $S_i$ individuals can be infected not only by individuals from compartment $I_i$ but from all compartment $I_g$ at all age groups, $g\in\{1,\dots,n\}$. Therefore, change in any infectious compartment affects the propensity for transitions in all age groups. Thus upon an infection event all propensities $a_{SE}^g, g\in\{1,\dots,n\}$ have to be updated. In more complicated models there may be several infectious classes (such as latent or in case of ebola models even dead individuals may cause infection). To this end, in general, we flag (*) the classes $c^*$ that influence propensities across all age groups and also flag transitions $t^*$ that have to be updated for all age group whenever a compartment from a flagged class changed. We handle this issue in the computer implementation. 

\begin{figure} 
	\centering
	\begin{minipage}{0.45\linewidth}
        \includegraphics[clip,trim=20 0 20 0,scale=0.4]{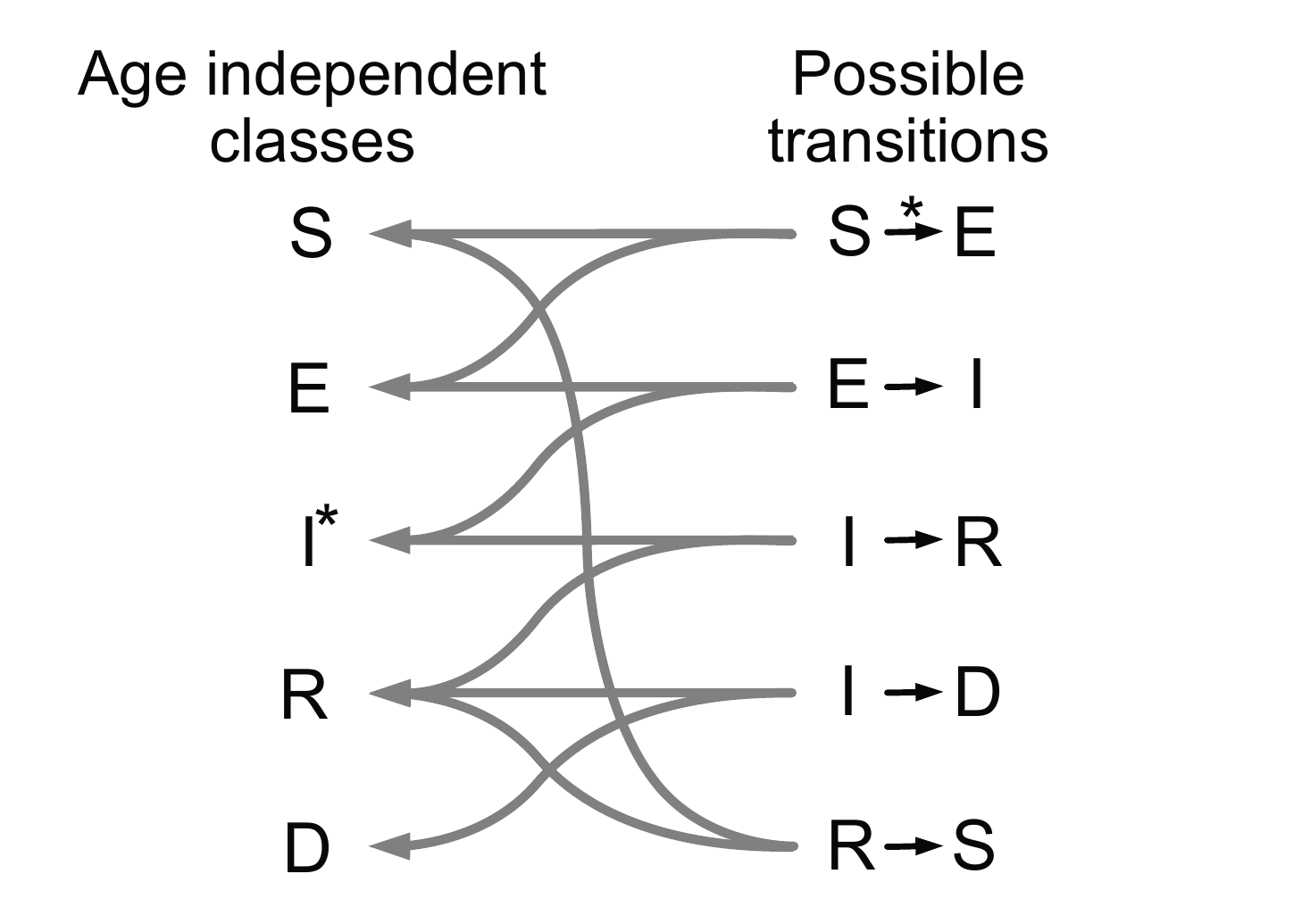}
	\end{minipage}\begin{minipage}{0.45\linewidth}
		\centering
		\includegraphics[clip,trim=20 0 20 0,scale=0.4]{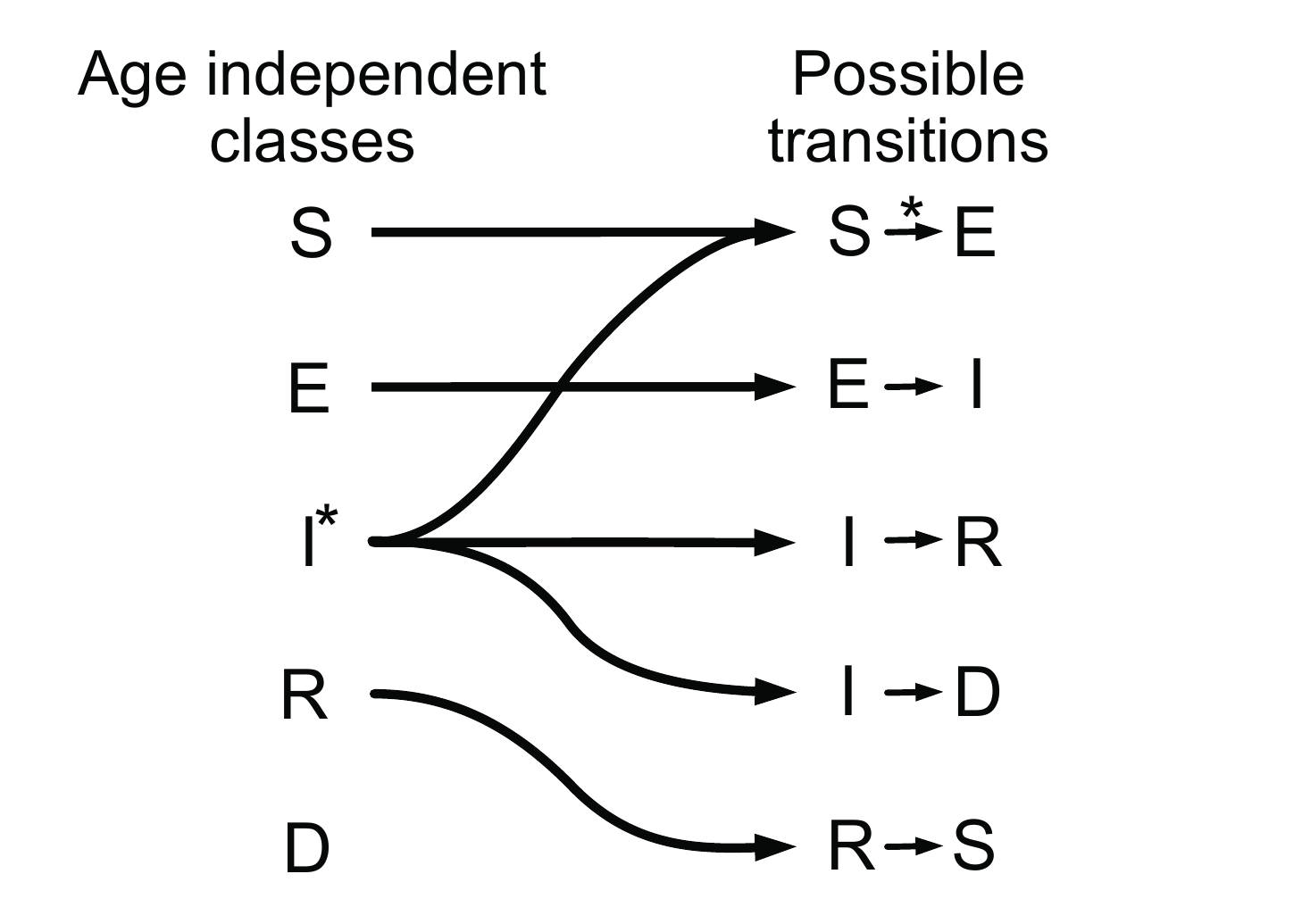}
	\end{minipage}  
	\caption{\textbf{Update graph of the SEIRD model} The figure shows the bipartite update graph that contains information about the classes and propensity functions to be updated upon event selection.}\label{fig:UG}
\end{figure}

For instance the two part of the UG of the SEIRD model with waning immunity is shown on Fig.\ref{fig:UG}. The left part of the figure shows edges $t_{X_jX_k}\rightarrow c$ and the right part shows edges $c\rightarrow t_{X_jX_k}$. Class $I^*$ and transition $S\xrightarrow[]{*} E$ is flagged thus whenever an $I^i$ compartment changes all propensities $a_{SE}^i$ need to be updated.

During the simulation, after the event selection step we end up with the coordinate triple $(i,j,k)$ that selects transition $t_{X_j,X_k}$ at age group $i$. By using the $t_{X_j,X_k}\rightarrow c$ edges of the graph (left figure) we update the state variables according to $X_j^i(t+\tau)=X_j^i(t)-1$, and $X_k^i(t+\tau)=X_k^i(t)+1$ and all other $X_f^i(t+\tau)=X_f^i(t)$ remain unchanged. Next we update the propensity $a_{jk}^i$ according to the $c\rightarrow t_{X_j,X_k}$ edges. If we encounter and edge that points from a flagged class to a flagged transition then we updated all $a_{jk}^i$ for $i\in \{1,\dots,n\}$.

\subsection{Propensity matrix method extended with the update graph}

\begin{enumerate}
	\setcounter{enumi}{-1}
	\item \textbf{Initialisation step:}
	\begin{itemize}
	    \item Fix the order of the state variables: $(X_1^1,\ldots,X_l^n)$
	    \item Construct the Update Graph according to Sec. \ref{sec:update}
	    \item Set $t\leftarrow 0$, initial values $(X_1^{1}(0),\ldots,X_l^{i}(0))$, and halting condition $C_H$.
	    \item \textbf{Calculate propensity matrices} $M_i$ for all age groups $i\in\{1,\ldots,n\}$
	\end{itemize}
	
	\item \textbf{Selection step:}
	\begin{itemize}
	    \item \textbf{Choose intervent time} from $P(\tau)=a\cdot e^{-\tau\cdot a}$ where $a$ comes from \eqref{eq:a},
	    \item \textbf{Choose the next reaction} $t_{X_j,X_k}$ at age group $i$ from the distribution  $p_{j,k}^i=\frac{a_{j,k}^i}{a}$ \eqref{eq:prop_mtx_event},
	\end{itemize}
	
	\item \textbf{Update step:}
	\begin{itemize}
	    \item \textbf{Time:} $t\leftarrow t+\tau$
	    \item \textbf{Update state variables:} by using the $t_{X_j,X_k}\rightarrow c$ edges of the graph update $X_j^i(t+\tau)=X_j^i(t)-1$, and $X_k^i(t+\tau)=X_k^i(t)+1$ and all other $X_f^i(t+\tau)=X_f^i(t)$ remain unchanged,
	    \item \textbf{Update propensities:} 
	    update the propensity $a_{jk}^i$ according to the $c\rightarrow t_{X_j,X_k}$ edges. If we encounter and edge that points from a flagged class to a flagged transition then we updated all $a_{jk}^i$ for $i\in \{1,\dots,n\}$.
	\end{itemize}
	
	\item \textbf{Halt} if $C_H=True$ else \textbf{continue} with step 2.
\end{enumerate}

\section{Experiments}\label{experiment}

In this section, we demonstrate the advantages of the stochastic approach and the flexibility of the propensity matrix - update graph method by experimenting with some real life problems.

\begin{figure}
	\centering
	\includegraphics[width=0.8\linewidth]{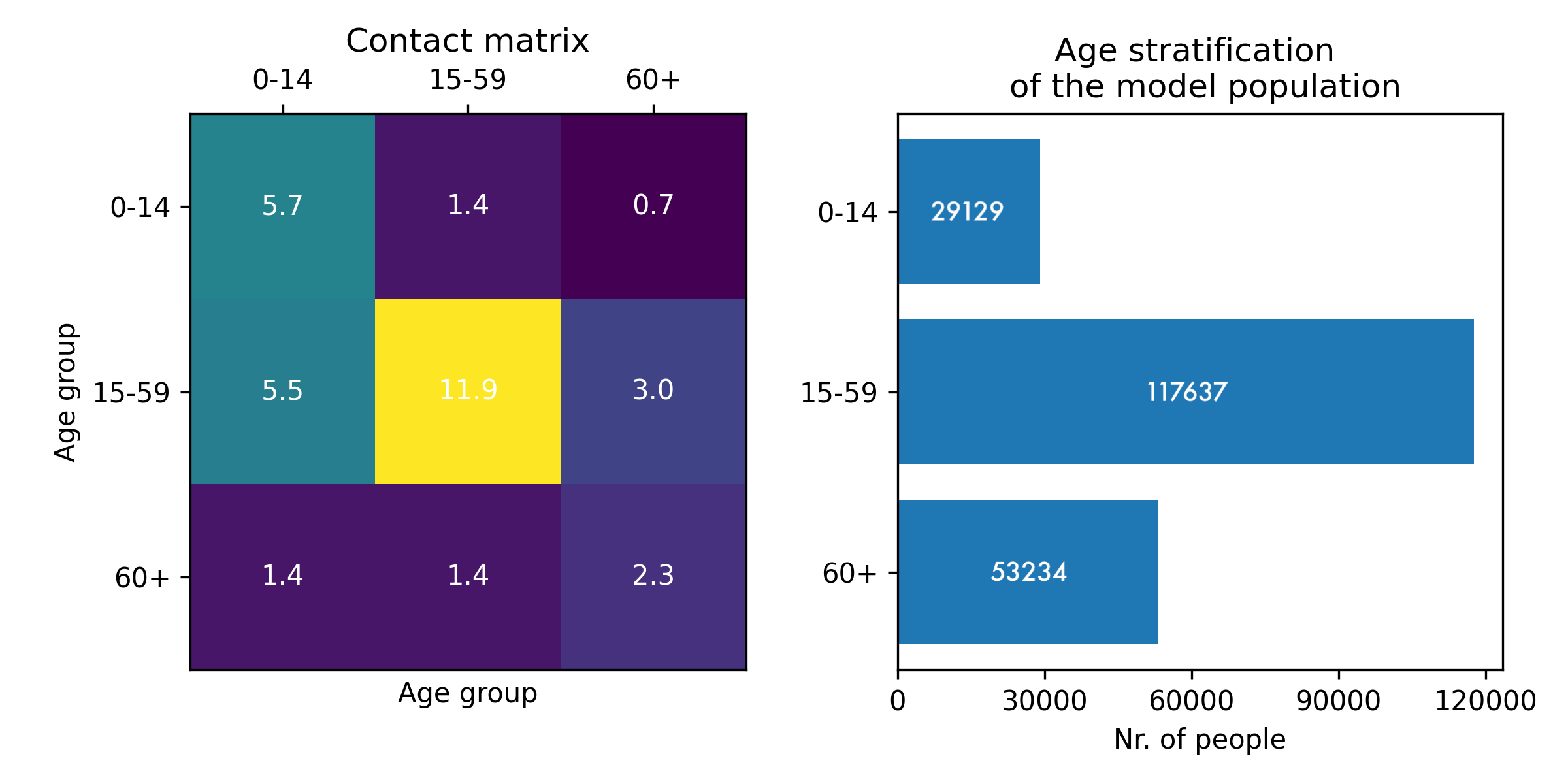}
	\caption{\textbf{Population related parameters of the demonstrated experiments}\newline
	Agregated contact matrix (left) and age stratification (right) of the investigated model population. Both data collection corresponds to the Hungarian society.}\label{fig:population}
\end{figure}

During the following simulations, we consider a population of 200,000 individuals divided into three age groups (0-14, 15-59 and 60+), with the aggregated contact matrix from Prem et al.\cite{Prem} and with the population distribution of Hungary from the Hungarian Central Statistical Office (KSH) age-stratification\cite{ksh} (cf. Figure \ref{fig:population}). We symmetrised the contact matrix according to Sec. 2.3.3. in Röst et al.\cite{Rost}.

Using the stochastic SEIRD model with immunity waning from Sec.\ref{sec:SEIRD} we investigate the course of a COVID-19 epidemic with aggregated parameters from Rost et al.\cite{Rost} (cf. Table \ref{tab:params}). Every simulation starts with $I_1(0)=10$ infected individuals in age group 1 at $t=0$ -- all other individuals considered to be susceptible according to the age partition ($S_1(0)=29119, S_2(0)=117637,S_3(0)=53234$). The governing propensity functions can be found in Sec. \ref{sec:SEIRD}.

\begin{table}
\tbl{Parameters of the demonstrated epidemic}
{\footnotesize
    \begin{tabular}{@{}lc@{}}
        \hline
        {} &{}\\[-1.5ex]
        Parameter & Value/age-dependent vector\\[1ex]
        \hline
        {} &{}\\[-1.5ex]
        Incubation period ($\epsilon^{-1}$)& $5.2$ (days)\\[1ex]
        Infectious period ($\gamma^{-1}$)&$5.0$ (days)\\[1ex]
        Time spent immunized ($\omega^{-1}$)&$180$ (days)\\[1ex]
        \hline
        {} &{}\\[-1.5ex]
        Infection rate ($\beta$) & $0.05$\\[1ex]
        Probabilty of death ($p_i$)&($0.0000451$, $0.00117$, $0.0281$)\\[1ex]
    \end{tabular}\label{tab:params}
}
\end{table}

\begin{figure}
	\centering
	\includegraphics[scale=.37]{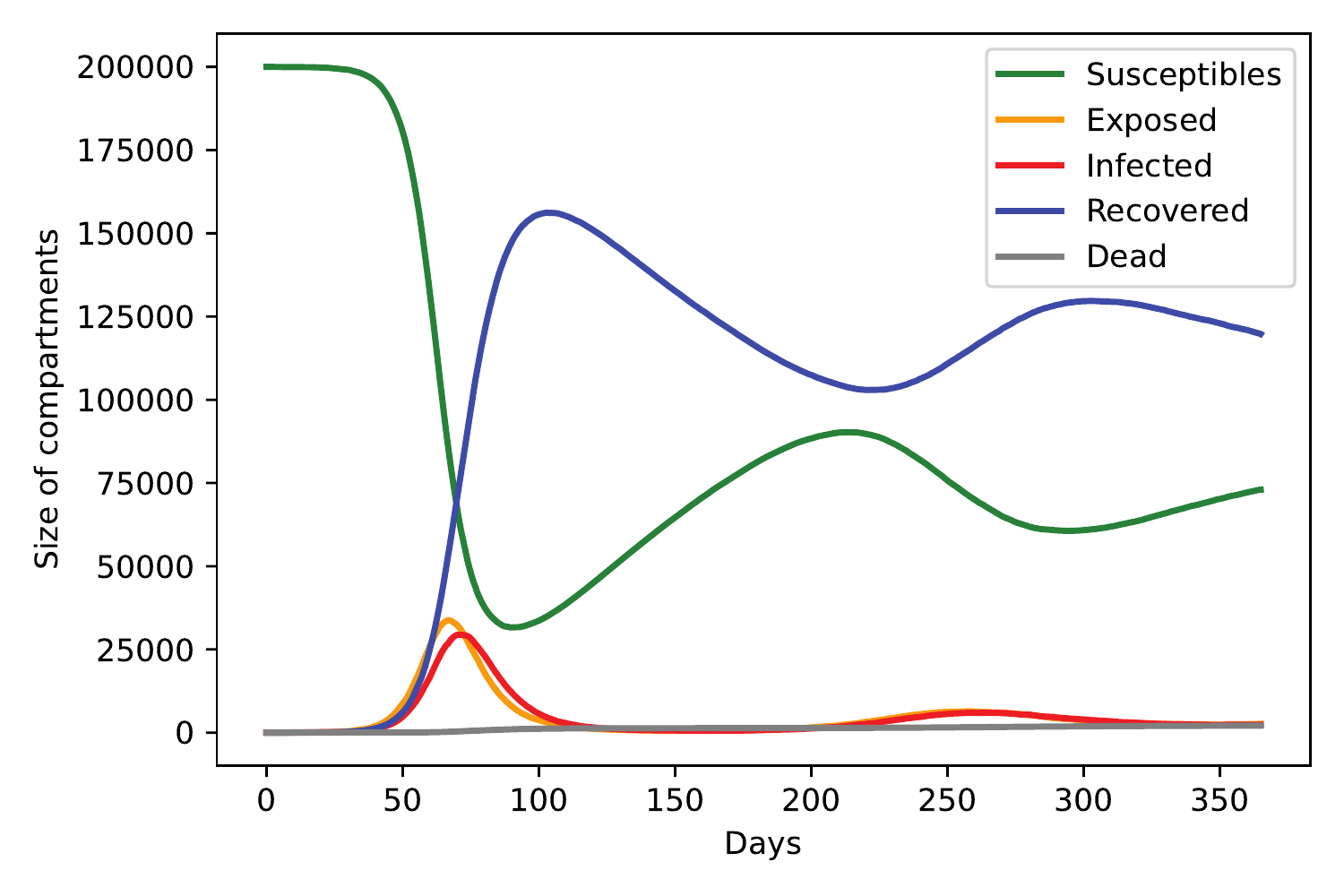}
	\includegraphics[scale=.37]{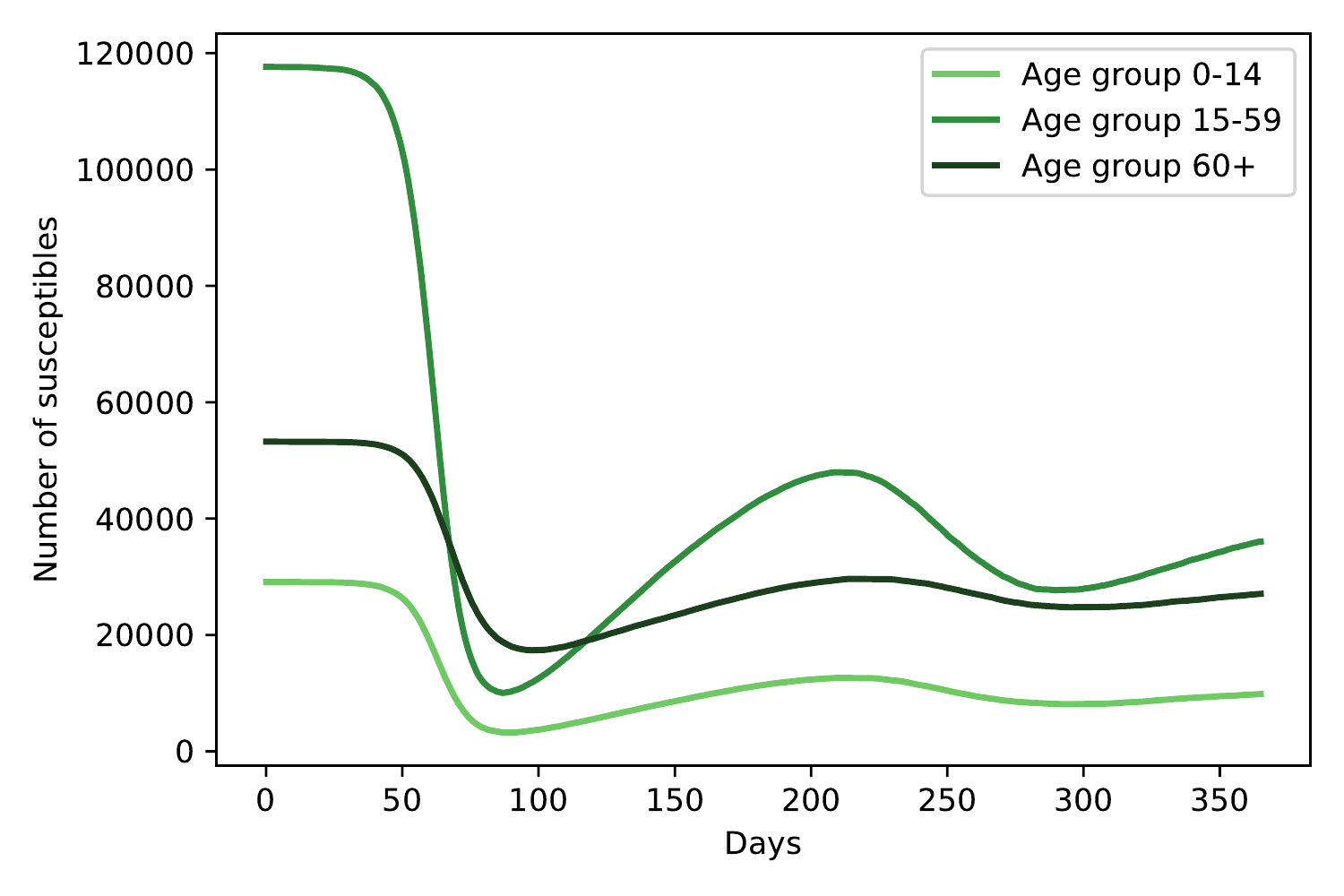}
	\caption{\textbf{Stochastic realisation of the SEIRD model with immunity waning}\newline 
	Time series of the age agregated state variables (left), and the three susceptible age groups (right) from the same realisation of the stochastic SEIRD model. For the parameters cf. Fig.\ref{fig:population} and Tab.\ref{tab:params}. \label{fig:compartments}}
\end{figure}

The left part of figure \ref{fig:compartments} shows the time series of the age aggregated state variables $\sum_{i=1}^3 S_i,\sum_{i=1}^3E_i$, etc. from a single simulation outcome. We can observe that during the "first wave" the exposed class peaks before the infected class, and class $S$ and $R$ shows oscillation due to loss of immunity. The right part of Figure~\ref{fig:compartments} shows the time series of the three age groups of class $S$ from the same simulation. 

\begin{figure}[h!]
	\centering
	\includegraphics[scale=.4]{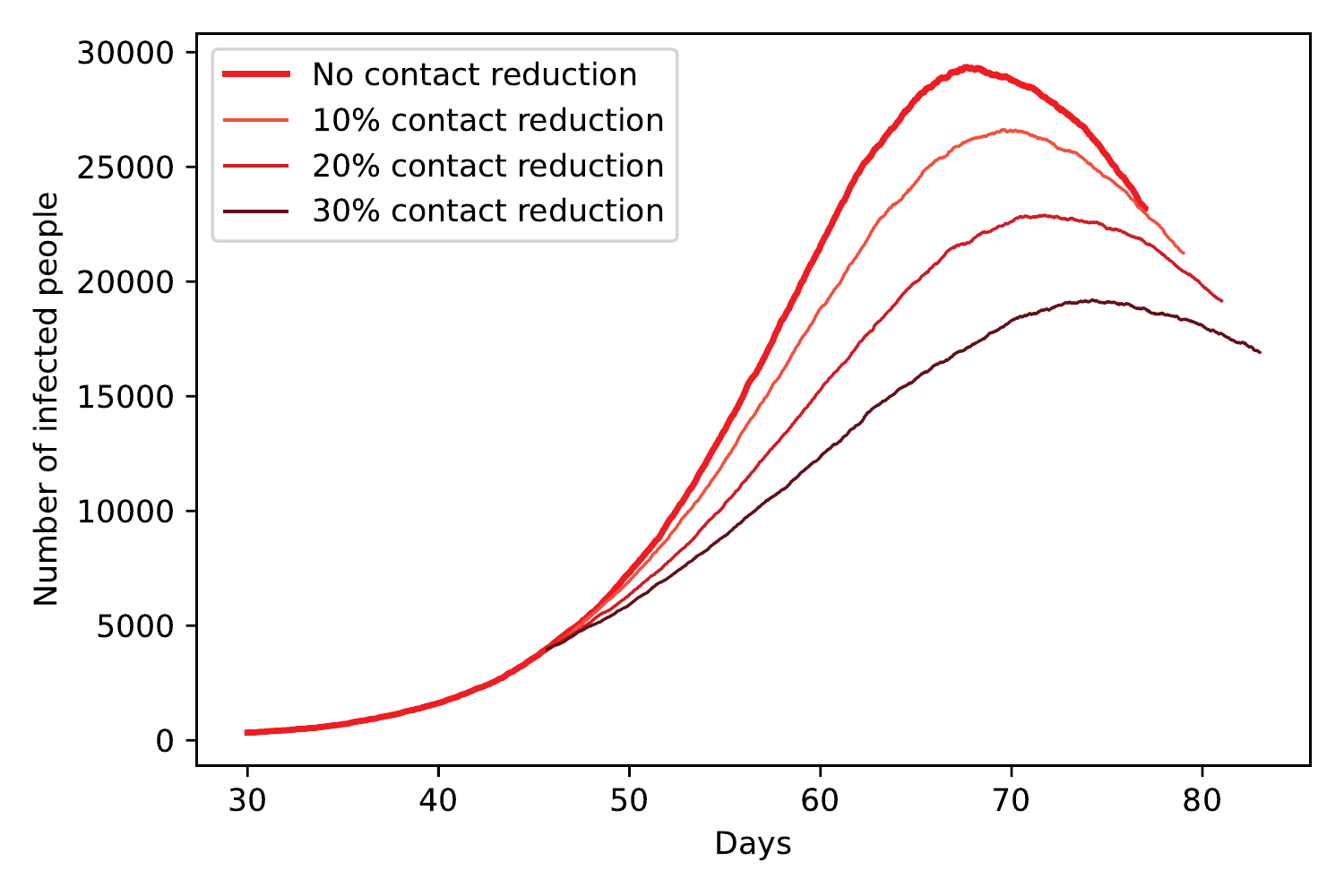}
	\caption{\textbf{Effect of contact reduction in the early stage of the epidemic}\newline
	In this simulation we applied a $10\%, 20\%$ and $30\%$ uniform contact reduction in the total contact structure of the population. The simulation stops 10 days after the peak. \label{fig:multiple}}
\end{figure}

On Figure~\ref{fig:multiple} we focus on the peak size of the epidemic and show the effect of contact reduction in the early stage of the disease spread (on day 45 in this case). By $10\%, 20\%$ and $30\%$ of contact reduction we mean that we multiply every element of the contact matrix with $0.9, 0.8$ and $0.7$ respectively. The simulation was halted 10 days after the peak. We can notice that a uniform contact reduction of $30\%$ in every age group may decrease the peak size by $1/3$, and in the meanwhile it also delays the time of the peak.

Montecarlo simulations, such as any version of Gillespie's SSA can only serve information about the process if we run several stochastic realisations and investigate the statistics of them. Thus it is always important to know how many simulations we need for plausible conclusions.

In the following experiments we run a large number of simulations the following way: we obtain 100 simulations (for having enough data for statistical analysis) and for the daily sampled time series of every single compartment $X^i_{j}(t)$ we evaluate the mean $\mu_{X_j^i}(t)$ and variance $\sigma_{X_j^t}(t)$.
We also calculate the mean value and variance for the time of the peak $(\mu_t,\sigma_t)$ and for the size of the peak $(\mu_s,\sigma_s)$. Then after each simulation we check whether 95\% of the collected first peak sizes are in the confidence interval $[\mu_s-2\sigma_s, \mu_s+2\sigma_s]$ as well as in $[\mu_t-2\sigma_t, \mu_t+2\sigma_t]$ and stop the routine, if this condition is fulfilled.

\begin{figure}
	\centering
	\includegraphics[scale=.37]{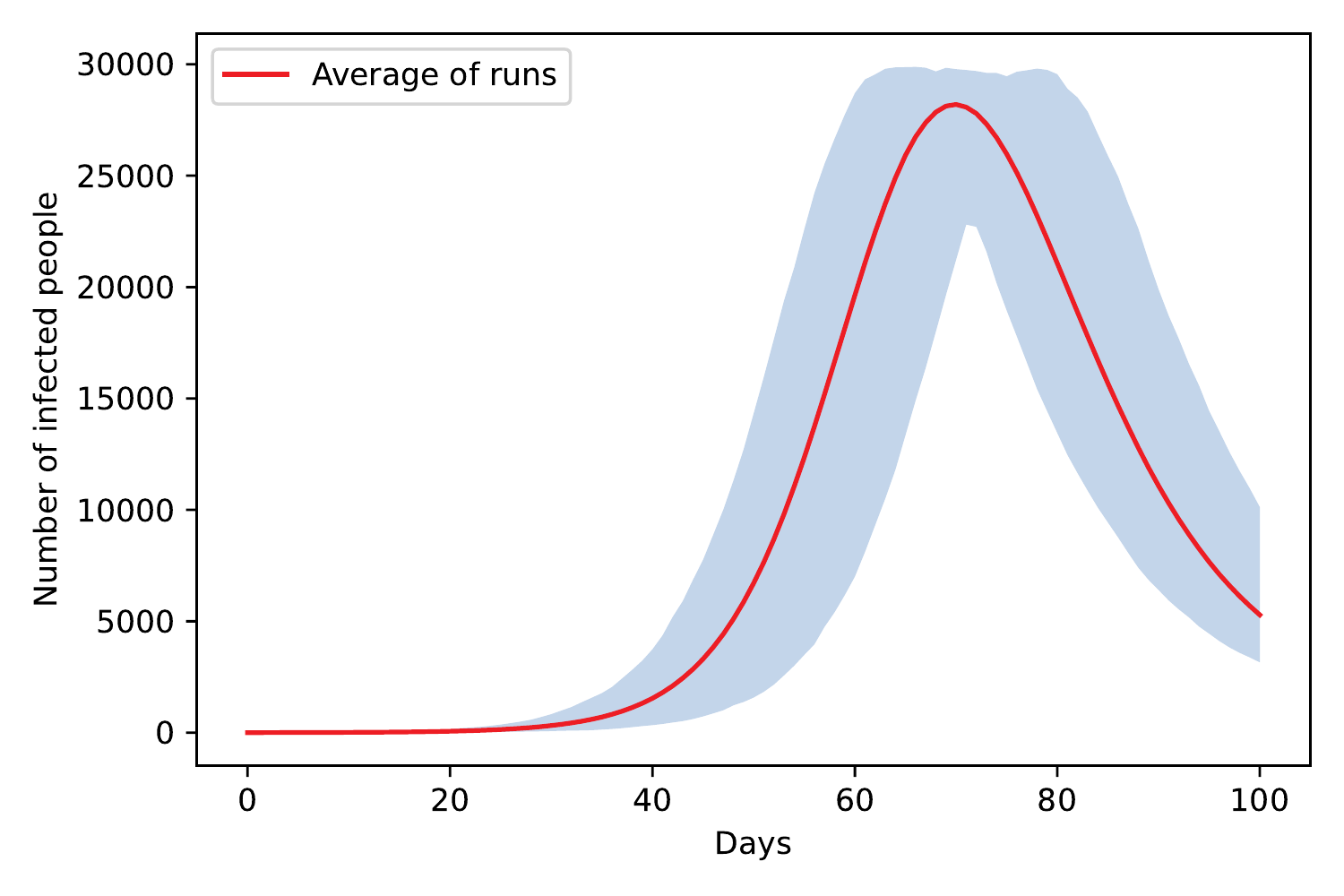}
	\includegraphics[scale=.37]{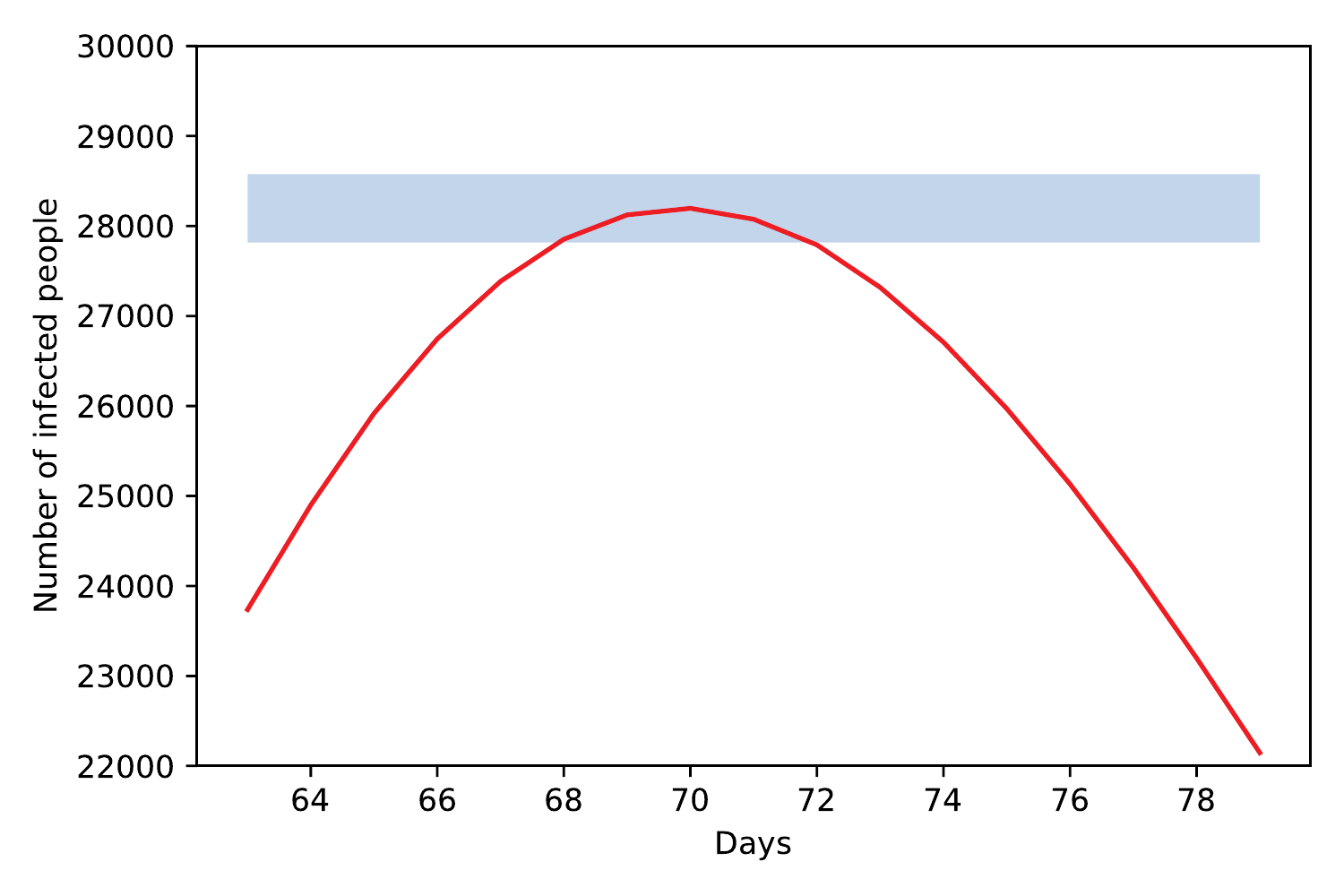}
	\caption{\textbf{Time series of the daily sampled aggregated infected classes}\newline
	The red curve shows the daily sampled  $\sum_{i=1}^3\mu_{I_i}(t)$ in case of 110 simulations around the peak of the first wave. The blue band on the left shows the daily sampled minimal and maximal values of the stochastic ensemble, the blue rectangle on the right represents the $2\sigma_t$ wide and $2\sigma_s$ high confidence interval around the maximum of the mean.
	\label{fig:confidence}}
\end{figure}

On both part of Fig.~\ref{fig:confidence}, with solid red curve, we plotted the mean of the daily sampled time aggregated infected compartments $\sum_{i=1}^3\mu_{I_i}(t)$. 
The blue band on the left plot shows the minimal and maximal values we calculated for each days from the ensemble of the independent realisations of the process as well. Let us remark that the band at the top (around the peak) is almost flat, meaning that the different simulations produced similar peak sizes but different peak times.
On the right part of Fig.~\ref{fig:confidence} the highlighted rectangle shows the $2\sigma_t$ wide and $2\sigma_s$ high confidence interval around the maximum of the mean. Thus we may conclude that the peak of the epidemic occurs between day 63 and 79 and it is expected to be between approximately 27,900 and 28,600 with 95\% probability.

\section{Summary and discussion}

In this paper we introduced our approach that we use in our forecasting work to efficiently implement age structured stochastic epidemic models. The core idea, the so called propensity matrix method, serves a data structure to handle the propensities and select the upcoming reaction/event to execute in a convenient way, in case of robust epidemic models. In Section~2 we introduced this method and the algorithm that obtains the time evolution of the state variables. In Section~3 we further improved the algorithm by introducing the so called update graph that helps us to speed up the algorithm by updating only the minimal number of propensities that are required to be updated due to the change in the state variables. Finally in Section~4 we showed some real life experiments to demonstrate the strength of the stochastic approach and the flexibility of our method.

The speed of the simulation is usually a limiting factor of stochastic simulations. Generations of single trajectorie requires several independent runs and therefore are typically expensive in terms of computational time. From coding perspective interpreted languages like Python may have low performance. The simulation time may be reduced by several magnitudes using low-level languages like C or C++. The propensity matrix algorithm can be easily applied to such languages. We provide our sample code in Python to emphasize the algorithm and help readability as much as possible.

However, from the modeling perspective simulation of whole trajectories during years, from the outbreak untill the very end of the epidemic, is usually unnecessary in practice. Information provided by stochastic simulations is particularly valuable i) in the beginning of the epidemic - when only a small portion of the population is infected, ii) in case of parameter estimation, iii) when we want to make short-term forecasts, or iv) we want to gain information of the variance of state variables in a short time scale - such like in Section \ref{experiment} where we estimated the number of infected individuals near the peak of the epidemic. Thus, reducing the scope of the simulation to shorter time intervals is usually a convenient way of dealing with speed.

Incorporating demography and aging into our models may change the process drastically in a long term and may lead to much more realistic models. However, in case of short term simulations it usually leads to unnecessary complications and a great number of (aging) events that occupies the simulation engine and it leads to slower simulations. Moreover in case of a small number of age groups the change between two age groups during a short term simulation remains negligible. However, if unavoidable, we suggest that instead of Gillespie's\cite{Gillespie-1} exact stochastic algorithm consider an approximating algorithm when aging events occur in discrete time steps, say on every 30th day of the simulation, and the appropriate portion of the population in compartment $X_j^i$ (for $j\in\{0,\dots,l\}$) moves to compartment $X_j^{i+1}$ if $i\neq n$ or dies whenever $i=n$. In the meanwhile appropriate portion of newborns need to be added into the susceptible, immunised or the infected compartments of age group 1.

\end{document}